\newlength{\bxwidth}\bxwidth=3.2 truein
\newcommand\om { \omega}

\newlength{\jight}
\newlength{\fight}
\newcommand\ltdash{\raise-1.8pt\hbox{$\scriptscriptstyle |$}}
\newcommand \beq  {\begin{equation}}
\newcommand \eeq  {\end{equation}}
\newcommand \bea {\begin{eqnarray} }
\newcommand \eea {\end{eqnarray}}

\documentclass{LT23auth}
\usepackage{graphicx}

\begin{document}

\begin{frontmatter}

% Use lower case letters in the title.
\title{Breakdown of the Fermi liquid theory in heavy fermion compounds}

\author[address1]{C. P{\'e}pin\thanksref{thank1}}
\and
\author[address2]{P. Coleman},

\address[address1]{SPhT,CEA-Saclay, l'Orme des Merisiers,
91191 Gif-sur-Yvette, France.}

\address[address2]{Serin Laboratory, Rutgers University,
P.O. Box 849, Piscataway, NJ 08855-0849, USA.}

% The corresponding author should be distinguished and his email
% address and/or fax number must be given. His mailing address has to
% be complete: the proofs are send to this address around
% January 1, 2003. The address for sending proofs has to be indicated
% as "present address", if it is different from the address above.
\thanks[thank1]{Corresponding author.
 E-mail: pepin@spht.saclay.cea.fr}

\begin{abstract}
We review the anomalous properties of heavy fermion compounds like
CeCu$_{6-x}$Au$_x$ or CeMIn$_5$ close to a zero temperature phase
transition called a quantum critical point. Anomalous behavior of the
resistivity, specific heat and magnetic properties observed in various
compounds
%suggest
suggests
a fundamental breakdown of the Fermi liquid
theory. Recent measurements on YbRh$_2$Si$_2$, field-tuned through the
quantum critical point, provide interesting insights on the evolution
of the Fermi liquid close to criticality.
%All this body of evidence
%suggests some kind of spin and charge separation at the quantum
%critical point.
We discuss the possibility of a kind of spin-charge separation at the
quantum critical point and make some remarks about its possible link
with the underscreened Kondo model.

\end{abstract}

\end{frontmatter}

\section{Experimental overview and theoretical insights}
\subsection{experiments}
%% Landau Fermi liquid theory is not a notion, but a concept.
%% Why talk about when it was done, unless you use this information
%% in the text- it must be made relevant.
%The notion of Fermi liquid (FL) theory~\cite{landau} was derived
Fermi liquid  theory~\cite{landau}, developed by Landau
%%in the 1960's
in the mid 1950's, provides the foundation for much
of our current understanding of electron fluids.
According to the Landau  Fermi Liquid theory (LFL),
thermodynamic and transport properties of
a metal at very low temperatures
%%can be described in terms of
are described in terms of weakly interacting fermions, or
``Landau quasiparticles'' .
%% Uneccessary extra article
%% The
Landau quasiparticles
%% a piece of music is composed by someone, consists, or ``composed
%% of'' would be better here
%% composed by
consist
of electrons,
surrounded by a cloud of spin and charge
polarization. They share the same quantum numbers as free electrons
but their masses can be strongly renormalized
%% through
by the back
flow of
%% many body interactions.
the surrounding fluid.
%% Better not to say something is the ``most'' , be modest.
%%The most
A striking example of the
robustness of the LFL is
provided by
%%the (no article needed)
heavy fermion behavior, where the
effective mass of the quasiparticles
%reaches a
can be hundreds of times  greater
than that
%%the one
of a bare electron. The physical properties of this fluid
%%though, - why though?
follow
%% self referential- needs to be changed
%% the
certain characteristic power-law dependences
%%of the FL
%%with
on temperature.
For example,
the specific heat has a linear temperature
dependence, the resistivity a quadratic one.

%% Recently, a
A wide  body of experimental results
%% I think one should refer to the compounds separately
%%on compounds like
%% CeCu$-{6-x}$Au$_x$~\cite{hvl,schroder} or on
%5 CeMIn$_5$~\cite{sarrao}
%% not  suggest- needs to be stronger
% suggest
show that when heavy fermion
materials are
tuned to a zero temperature antiferromagnetic transition,
their physical
properties
%%show a strong departure
depart qualitatively from
%FL
LFL
theory.
There is a growing list of heavy electron materials that
can be tuned into the quantum critical
point by, either
alloying, such as $CeCu_{6-x}Au_x$\cite{hvl} or the layered materials
$CeMIn_5$\cite{sarrao}, through
the direct application of pressure, as in
case of $CeIn_3$\cite{mathur} or $CePd_2Si_2$\cite{grosche}, or
via the application
of a magnetic field, as in the case of $YbRh_2Si_2$\cite{gegenwartce}.
At the quantum
critical point (QCP) the key properties suggesting a breakdown of
the Fermi liquid are:
\\

\begin{itemize}
\item the specific heat coefficients diverge at the
QCP\cite{steglichmass,aoki,schroder}. For most cases the
divergence displays a logarithmic temperature dependence
 \bea
 \gamma(T) =   \frac{C_v}{T} = \gamma_0 \log[ \frac{T_0}{T} ] \ .
 \eea This implies that the
%%a divergence of
the quasi-particle
 mass diverges and the Fermi temperature goes to zero
%%or in other words, a renormalization of the Fermi
at a QCP.
 \beq \begin{array}{cc}
 \frac{m^*}{m} \rightarrow \infty \; \; \; \; \; \; \; \; \; & T_F^* \rightarrow 0
 \end{array} \ .
 \eeq

 \item a quasi-linear temperature dependence of the resistivity,
 contrasting with the quadratic dependence of the
 FL~\cite{grosche,gegenwartce,julian}.
 \beq \rho \propto T^{1+ \epsilon} \eeq with $\epsilon$ in the
 range of $0-0.6$. Several compounds including
 YbRh$_2$Si$_2$~\cite{trovarelli} and
 CeCu$_{6-x}$Au$_x$~\cite{schroder}
 have perfectly linear resistivity, a property
 reminiscent of the normal phase of high temperature
 superconductors.

 \item the spin susceptibilities acquire anomalous exponents
 \beq \chi^{-1}(T) = \chi_0^{-1} + T^a \eeq
 with $a <1$ for CeCu$_{5.9}$Au$_{0.1}$,
 YbRh$_2$(Si$_{1-x}$Ge$_x$)$_2$ ($x=0.05$) and CeNi$_2$Ge$_2$\cite{grosche}.
In the case of CeCu$_{6-x}$Au$_x$~\cite{schroder}
%%enough  (not needed)
 magnetization measurements reveal that at finite
fields $H$, the differential magnetic susceptibility
exhibits exhibits $H/T$ scaling.
%% Catherine- H/T was only measured in the static susceptibility
%%and $H/T$,
Neutron scattering measurements on the same material
\cite{schroder} reveal the presence of $\omega/T$
scaling in the dynamic
%% magnetic - it is ``dynamic spin susceptibility''
spin sysceptibility,
which may be cast into the form
 \begin{equation}\label{}
 \chi^{-1}( {\bf q},\omega ) = f({\bf q}) + ( i \omega + T )^a.
 \end{equation}
where the function $ f({\bf q})$ vanishes in the vicinity
%% Catherine- be very careful- where f (q) vanishes is
%% quite complex- a line, or butterfly shaped region
%% in momentum space.
%%at the
of the ordering wave vector. The $q-$ independence of second term
suggests a local origin to the damping of the
critical spin fluctuations, an observation that has stimulated
recent efforts to develop a self-consistent locally quantum-critical
model of the heavy electron QCP.\cite{qmsi}
Remarkably, a single exponent $a\sim 0.75$
governs both the $H/T$ and $\omega/T $ scaling.
The presence of
 the anomalous exponent $a$ suggests that  the interaction amongst
the quantum critical modes has renormalized to
%%the system
%%is at
strong coupling. This fact is made clear by the scaling laws in
 $\omega/T$ and $H/T$.
%One can show that
This type of ``naive''
 scaling property (quantities scale according to their dimensions)
 is the hallmark of a system lying below its upper critical
 dimension.\cite{ourreview}

 \end{itemize}

\subsection{Key theoretical insights}
%% %%
%% This sentence does not
%% make sense to me at all.
%%
%%The experimental observations at criticality underline the concept
%%of universality
The experimental observations at criticality suggest
an underlying universality in the physics.
~\cite{ourreview}.
%% These next two sentences need to be brought together.
%%The effective Fermi temperature
%%$T_{F*}$ is renormalized to zero at the QCP, suggesting that
%%temperature is the unique energy scale
%which drives
%%the system
%to
%%criticality.
%%The fact that the dynamical susceptibility of
%%CeCu$_{5.9}$Au$_{0.1}$ can be expressed in terms of a scaling
%%function $F \left( \om / T, B / T \right)$ implies scale
%%invariance at criticality.
The ability to express the dynamical susceptibility of
Cu$_{5.9}$Au$_{0.1}$ in terms of a scaling function
$F \left( \om / T, B / T \right)$ and the
observation
that the effective Fermi temperature $T_{F}^{*}$renormalizes
to zero at the QCP, both suggest that underlying physics
at the QCP is scale invariant.
In this situation, we expect that
the physical properties will be independent of
the microscopic scales of the material- they are universal.

Moreover, the observation of anomalous exponents in the magnetic
susceptibility,
%as well as in
the transport and thermodynamic
properties are evidence
%for an
that the quantum critical physics may be described by an
effective theory whose coupling constants flow to strong
coupling. In particular the effective critical lagrangian must lie
below its upper critical dimension.
%%
%%  Catherine- what on earth is this stuff about causality
%%  linking the thermodynamics to the resistitivity. Please
%%  explain to me before putting it in!
%%
%%
%%We note also that transport
%%and thermodynamics measurements
%%are
%%tracking
%%with each other- the
%%logarithmic divergence of the specific heat coefficient can be
%%related by causality to the linear in T resistivity.

\subsection{Spin density wave scenarios}
\begin{figure}[hbt]
%h=here, t=top, b=bottom, p=separate figure page
\begin{center}\leavevmode
 \includegraphics[width=0.3\linewidth]{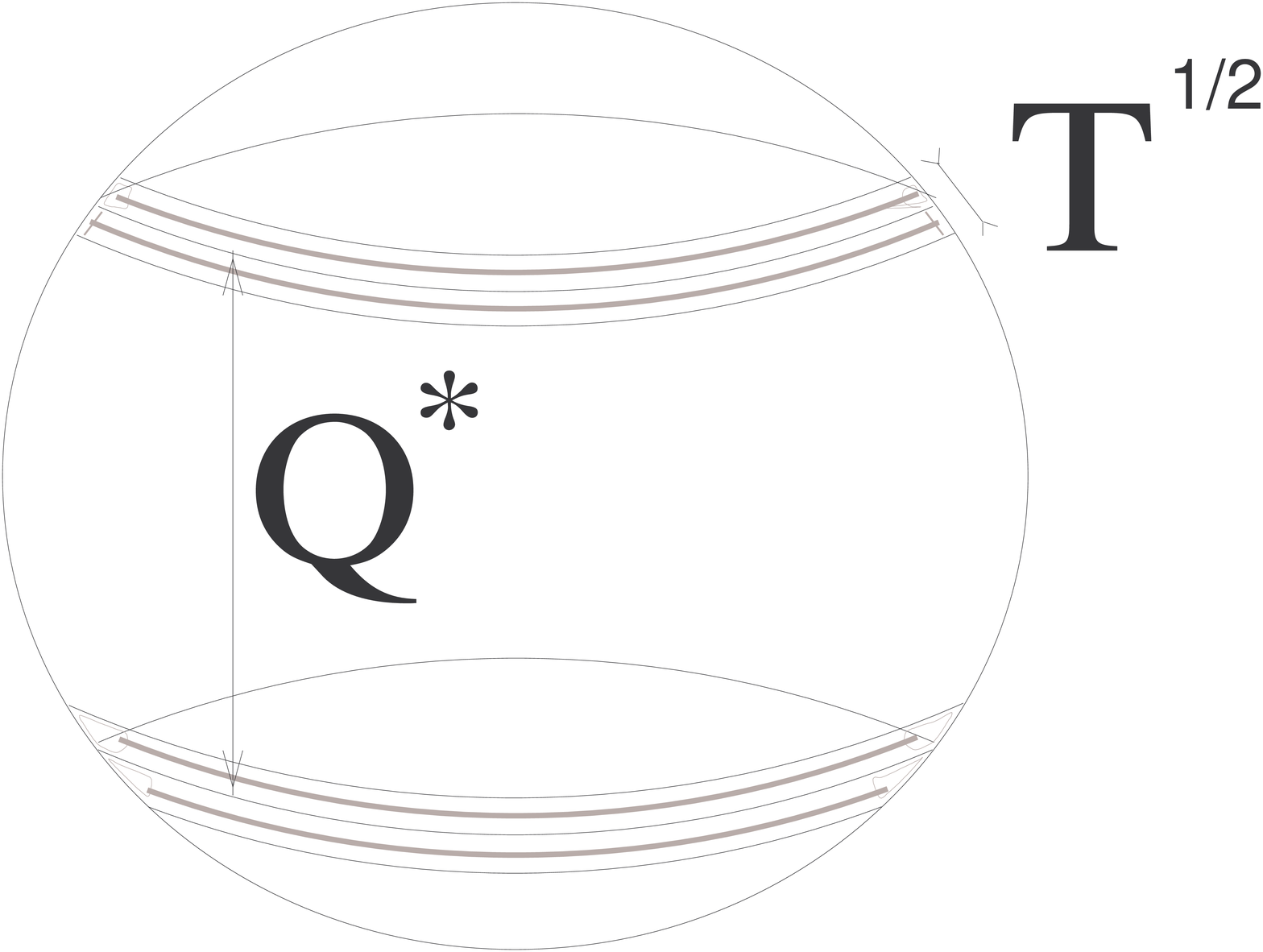}
  \hspace{10pt}
  \includegraphics[width=0.5\linewidth]{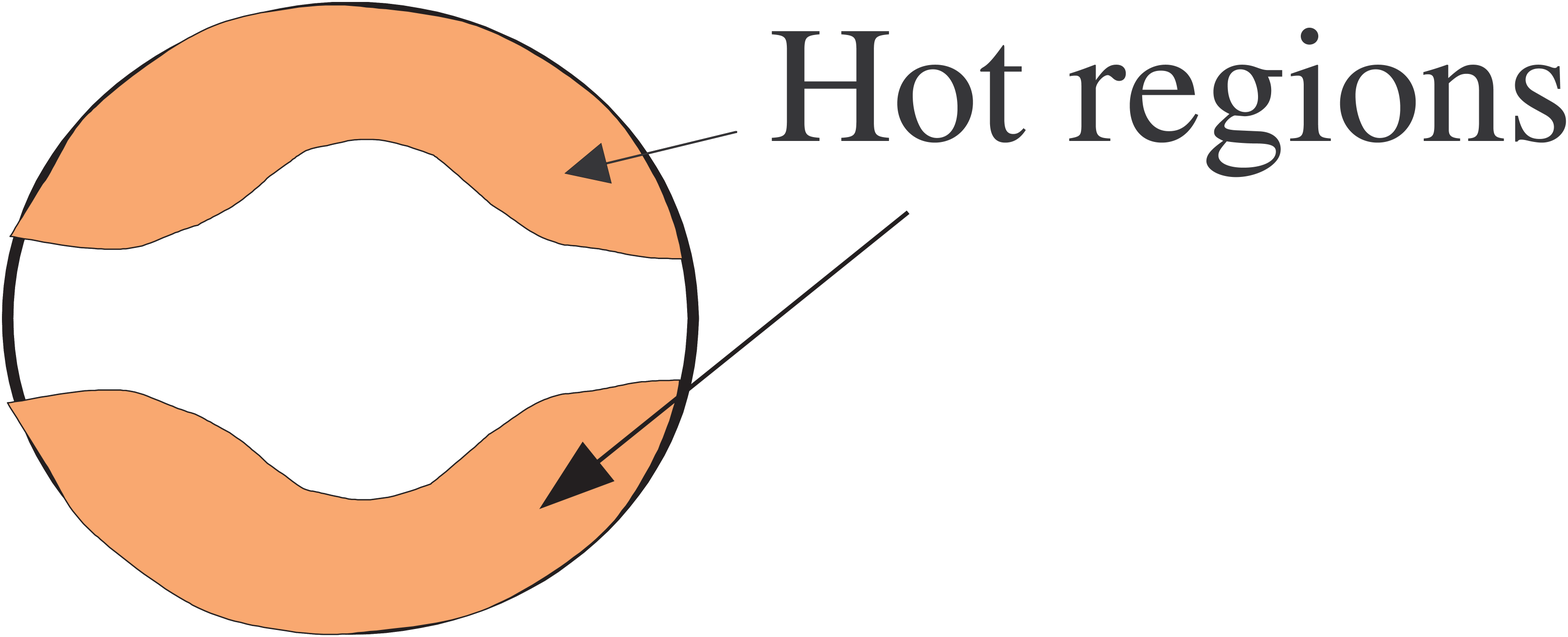}
  \caption{ Comparison between the 3D SDW Fermi surface topology
  (left), where hot lines are separated by the scattering wave vector
  $Q*$ and the 2D SDW scenario where a finite part of the Fermi
  surface is ``hot'' at $T = 0$. }
  \label{sdw}\end{center}\end{figure}

The constraints at criticality enable
%%US
us
to rule out the $3D$ spin
density wave (SDW) theoretical scenario~\cite{hertz,millis} for
the heavy electron quantum critical point (figure \ref{sdw},left).
In this scenario, magnetism develops by the spin polarization of
the Fermi surface and non-Fermi liquid (NFL) behavior results from
Bragg
%%scattering
%% diffraction  captures the idea of scattering off staggered
%% fluctuations
diffraction
of electrons off quantum critical spin fluctuations
in the magnetization. As shown in figure \ref{sdw}, three
dimensional spin fluctuations only couple strongly to electrons
along ``hot lines'' of the Fermi surface, separated by the wave
vector ${\bf Q}$ of the antiferromagnetic order.
%% In written English you do not begin a sentence with ``by''
%%By
The term
``hot lines'' refers to
%we mean
the ${\bf k}$ points on the Fermi surface
which are coupled the magnetic fluctuations at zero temperature.
The width of the hot lines increases with temperature like
$\sqrt(T)$. On the hot lines, the electrons are strongly perturbed
by the magnetic fluctuations and their life time is strongly
suppressed. More precisely in three dimensions the imaginary part
of the self-energy varies like $\Sigma_H \sim \omega$. Their
effective mass experiences a logarithmical increase with
temperature. Away from the hot lines however, the scattering of
the electrons off the critical modes is strongly suppressed - the
magnetic fluctuations getting massive. Within the 3D SDW scenario,
the hot lines are not wide enough to destabilize the FL theory.
%%
%% Catherine- the use of the ``we'' here can really upset readers
%% who actually did these calculations first.  I suggest:
%%We
In this situation,  the critical contributions
to the
free energy and the specific heat coefficient are given by
 \beq
\begin{array}{ccc} F \sim T^{5/2} \; \; \; \; \;
& \mbox{and} & \; \; \; \; \;  C/T = \gamma_0 -
 \sqrt{T} \end{array}
 \eeq
When the temperature is driven to
zero at criticality, the system
eventually
flows towards the FL
fixed point, keeping the universal exponents of the FL theory.
%\begin{figure} \centering \subfigure[]{
%  \label{figure1a}
%  \epsfig{figure=hotspot.eps,width=\jight}}
%\hspace{6pt} \subfigure[]{
%  \label{figure1b}
%  \epsfig{figure=2dsdw.eps,width=\fight}} \caption{ (a) Fermi surface with the ``hot''
%lines on of width $\Delta k \sim T^{ \eta}$ and (b) an
%illustration of the phase space restriction for the fermion
%scattering off magnetic modes, here for a ferromagnet. The angle
%between $\bk_2$ and $\bk_1$ is restricted by the spin fluctuation
%momentum $\bq$. }
%\end{figure}

In order to account for the properties at criticality, one can
imagine that the spin fluid is magnetically frustrated, dividing
into layers of decoupled plans. In this scenario, the 2D
SDW~\cite{rosch,paulkotliar}, the dimension of the spin fluid (2D)
is lower than the one of the conduction electrons (3D). Within
this assumption
%the whole or
a finite
%part
region
of the Fermi surface is
hot, as shown in figure \ref{sdw}, right. This intuitively
corresponds to an ``infinite'' width of the hot lines.
%%  It is not the assumption that destroys the Fermi liquid
%%  it is the 2D critical spin fluid.
%%This assumption-
This situation-
a fully anisotropic magnet-, while difficult to
realize experimentally, {\sl is} strong enough to destabilize the FL
theory, accounting for the logarithmic divergence of the specific
heat coefficient with temperature as well as the linear in $T$
resistivity.
Two
important remarks are due here. First, this scenario is remarkably
%unstable with respect
sensitive
to magnetic anisotropy.
If the 2D spin
fluid retains
%that
a small inter-plane
coupling, then Fermi liquid behavior will recover at temperature
scales smaller than the interlayer coupling.
%The 2D SDW scenario requires the spin
%fluctuations to be strictly two dimensional.
The second comment
relates to the upper critical dimension. Close to an
antiferromagnetic instability at zero temperature, the system
experiences strong fluctuations in time which scale with respect
to the length like
 $ \tau \sim \xi^z $.  $z$ is the dynamical
exponent of the QCP, and is equal to 2 for the antiferromagnet.
The effective dimension of the system is now $D=d+z$, where $d$ is
the number of spacial dimensions for the spin fluid.
In the 2D SDW scenario, $D=2+2= 4$ so the
effective lagrangian
%lays
lies at its upper critical dimension, implying marginal coupling
constants. This feature wouldn't be enough to account for the
anomalous exponents observed in the magnetic susceptibility.

%\begin{figure}[btp]
%h=here, t=top, b=bottom, p=separate figure page
%\begin{center}\leavevmode
%\includegraphics[width=0.8\linewidth]{figure1.eps}
%\caption{
%The size of the figure can be adjusted by changing the width number.
%Take care that the smallest characters in the figure are not much
%smaller than those in this figure caption.
%}\label{figurename}\end{center}\end{figure}

\section{A study of YbRh$_2$Si$_2$: the breakdown
of the 2D SDW scenario}

In the next section, we would like to
%%investigate in more details
discuss new insights into the
the mechanism by which the heavy quasiparticle
%electron THE ELECTRON NEVER DISINTEGRATES
may disintegrate at the QCP, revealed by
%% The compound is NOT FIELD tuned. The compound may BE field tuned TO
%% quantum criticality
recent studies
%  We focus on the study
of the compound
%%field tuned
  YbRh$_2$(Si$_{1-x}$Ge$_x$)$_2$ ~\cite{gegenwartfieldtuned,custers}.
  The undoped ($x=0$) compound lies remarkably close to
  a QCP, with a tiny antiferromagnetic ordering
  temperature $T_N = 70 mK$~\cite{trovarelli}
  that can be suppressed to zero by a
  small magnetic field of just $B_c = 0.06 T$ perpendicular to the
  easy axis. By allowing tiny amount of Germanium into the
  crystal, it is possible to fine-tune the N{\'e}el temperature
  $T_N$ and the critical field $B_c$ to zero, without appreciably
  increasing the disorder in the material. The materials can then
  be controllably driven back into the FL state by application of
  a small magnetic field.

  \subsection{The upturn in the specific heat}
\begin{figure}[hbt]
 %h=here, t=top, b=bottom, p=separate figure page
 \begin{center}\leavevmode
 \includegraphics[width=0.8\linewidth]{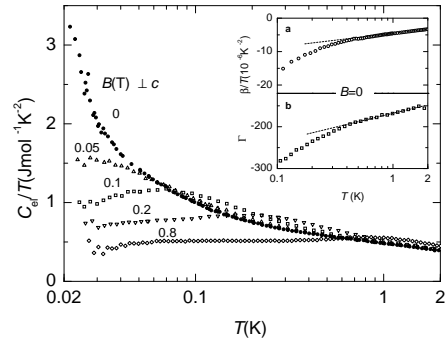}
  \caption{Strong upturn in the specific heat coefficient of
  YbRh$_2$(Si$_{1-x}$Ge$_x$)$_2$ for $x = 0.05$~\cite{custers}.
  One observes how
  the upturn fills in as one reaches the QCP by decreasing the applied
  magnetic field. }
  \label{gamma}\end{center}\end{figure}

   Several unusual features are revealed at criticality.
  In addition to the previously observed logarithmic
  increase when lowering the temperature,
  the specific heat coefficient $\gamma$ shows a strong upturn
  below $30 mK$ (figure \ref{gamma}). The electronic nature the upturn has been
  checked first by carefully removing the nuclear contribution,
  second by the observation that under a small magnetic field, the
  electronic contribution in the FL phase gradually saturates the
  upturns, as one gets closer to the QCP.
  Such an upturn of electronic nature reveals a strong departure
  from any SDW scenario, where the maximum possible increase of the
  specific heat coefficient is logarithmic.
  More specifically, the behavior of $\gamma$ can be cast into
  the form
  \beq \label{eqngamma}
  \gamma(T,B) = B^{-\alpha} \Phi ( \frac{T}{B^{\eta}} ) \eeq with
  $\alpha \sim 0.33$ and $\eta \sim 1$. $\Phi$ is a scaling
  function of the temperature and the magnetic field.
  The form of (\ref{eqngamma}) is remarkable in many respects. It
  reveals that two energy scales govern the behavior of the
  electrons in the FL phase. A first energy scale $T_F$ is
  associated to the inverse of the density of states.
  The second energy scale $T_0(B)$, characteristic of the
  excitations of the system appears in the scaling function
  $\Phi$. These two energies vary with the applied magnetic field
  as follows:
  \beq \begin{array}{ccc}
   T_F \sim B^{\alpha} \; \; \; & \mbox{and} & \; \; \; T_0(B)
  \sim B \end{array}
   \eeq  The scaling form (\ref{eqngamma}) of $\gamma$ ensures
  that at zero magnetic field
  $ \gamma(T,B=0) \propto T^{- \alpha / \eta} $. Since the
  exponent $\alpha$ is smaller than one, no residual entropy is
  sitting at the QCP.

%  Furthermore one can naively estimate what would be the effective
%  dynamical exponent $z$ which would account for
%  a divergence of the specific heat such as (\ref{eqngamma}).
  %%Indeed, the behavior of the entropy close to zero temperature is
  %%determined solely by the value of the dimension $d$ of the modes
  %%taken into account and the dynamical exponent $z$.
%By analogy with  a classical critical point, we might expect the
%singular part of the Free energy density $F_{sing}$ scales with
%the inverse of the correlation volume
%  in both space and time,
  % i.e.
%\begin{equation}\label{}
%F_{sing}\sim \frac{1}{\xi^{d}\tau }\sim \frac{1}{\tau ^{(1+d/z)}}\sim
%\left(\frac{k_{B}T}{\hbar } \right)^{(1+d/z)}.
%\end{equation}
%Here $\xi $ and $\tau \sim \xi^{z}$ are the correlation length and
%time respectively. At quantum criticality,
% temperature defines the
%temporal correlation length, $\tau \sim \hbar / ( k_{B}T)$, so
%yielding $F \sim T^{1 + d/z}$,
  %%leading to $C_v
  %%\sim T^{d/z}$
%so that  $ \gamma \sim T^{d/z -1} $.
%The condition $ d/z -1
%  \approx -1/3 $ implies $z\sim \frac{3}{2}d$. If the most singular
%contribution to the free energy derives from three dimensional
%fluctuations, then
%  $ z \approx 9/2 $, which would imply $\xi \sim T^{{-2/9}}$. This
%remarkably slow growth of the correlation length with temperature is perhaps
%further evidence for some kind of underlying locality to the
%physics. \cite{schroder,qmsi}

  \subsection{Kadowaki-Woods ratio is weakly field dependent}

 An alternative way of investigate the scattering behavior at the
QCP is to approach it from the FL phase, gradually reducing the
applied magnetic field towards its critical value $B_c$. Away from
criticality, the resistivity is quadratic $\Delta \rho (T) = \rho
(T) - \rho_0 = A(B) T^2 $ ($\rho_0$ is the residual resistivity),
below a characteristic temperature $T^*_F$. The temperature at
which the $T^2$ behavior, characteristic of the Landau FL theory
appears, sets the crossover between the heavy fermion regime and
the quantum critical regime. Inside the FL phase, even though the
quantum fluctuations are not strong enough to destabilize the FL,
the electron undergoes virtual scattering off the critical
magnetic modes. The variation of $A(B)$ close to the QCP provides
insight into these virtual excitations. Initial measurements on
germinium doped $YbRh_{2}Si_{2-x}Ge_{x}$ show that $ A(B) \sim 1/
(B-B_{c}) $ in the approach to the QCP. Over a wide range of
field, the Kadowaki Woods ratio is found to be approximately
constant, with a value
 \beq K= A/
(\gamma_0)^{2} \approx 5.8 \mu \Omega cm K^2 mol^2 /J
 \eeq is found to agree
within $ 40 \% $ with the empirical Kadowaki Woods ratio.
More recent measurements closer to the QCP show a slow upturn in
$K$ at very  small values of $B-B_{c}$.
\cite{custers}
Constancy of the Kadowaki woods ratio implies that the
transport scattering rate scales strictly with the
square of the renomalized
$T_F^* (B)$ of the heavy electron fluid. A truly field independent
 Kadowaki Woods ratio would indicate that the {\sl momentum} dependence
 of the scattering amplitude does not renormalize with
 $T_F^* (B)$.

\subsection{The 2D SDW scenario is ruled out}
 This set of data allows us to rule out the 2D SDW scenario. While
 the 2d SDW scenario leads to the result $A \propto 1/(B-B_c)$ as
 observed experimentally, the soft 2D spin fluctuations produce
 only a small renormalization in the heavy electron density
 of states with $\gamma \propto \ln 1/(B-B_c)$. This would give a
 strongly divergent Kadowaki Woods ratio in the approach to the
 QCP:
 $ K_{SDW} \propto 1 / \left ( (B-B_c) \ln^2 (B-B_c) \right ) $.
 The weak field dependence of the Kadowaki Woods ratio suggests
 that the most singular quasiparticle scattering amplitudes have a
 far weaker momentum dependence than expected in a spin density wave
 scenario.

 \section{The breakup of the heavy electron- is there a preformed
 Kondo effect at the QCP? }

%% The study of the YbRh$_2$(Si$_{1-0.05}$Ge$_{0.05}$)$_2$ compound has shed
%% light on a very unusual property. While the specific heat
%% coefficient experiences a strong upturn below $T = 30 mK $,
%% there is no reflection of this excess of entropy on the transport
%% properties- the resistivity stays linear in T for the whole
%% temperature regime. This suggests that at the QCP, some degrees
%% of freedom are present to enhance the entropy but don't
%% contribute to the transport. The picture which emerges is
%% reminiscent of the spin and charge separation in the cuprates
%% superconductors.
%%

 The study of YbRh$_2$(Si$_{2-x}$Ge$_{x}$) reveals
 a very unusual property. While the specific heat
 coefficient experiences a strong upturn below $T = 30 mK $,
%% there is no reflection of this excess of entropy on the transport
 there is no reflection of this sudden loss of entropy in the transport
 properties: the resistivity stays linear in T for the whole
 temperature regime.
%This suggests that at the QCP, some degrees
% of freedom are present to enhance the entropy but don't
% contribute to the transport.
This suggests that the degrees of freedom
associated with
the upturn in the specific heat are not involved
in the charge transport. What is the origin of this separation?
%%The picture which emerges is
%% reminiscent of the spin and charge separation in the cuprates
%% superconductors.

%% One can imagine that close to the antiferromagnetic transition,
%% the presence of very strong antiferromagnetic fluctuations
%% destabilize the formation of heavy fermion composite
%% quasi-particles. In other words at the QCP, the Kondo effect
%% leading to the formation of the composite electron is not
%% coherent, leaving behind some residual degrees of freedom
%% in the spirit of the under-screened Kondo problem. Of course here
%% the underlying lattice is of spins 1/2 and one would then naively
%% expect to form fully screened Kondo singlets like in the Fermi
%% liquid regime. Experiments suggest that
%% anti-ferromagnetic fluctuations somehow ``frustrate'' the Kondo
%% effect leading to a deconfinement of the composite electron at
%% the QCP.

%Although we still do not have a microscopic understanding of the
%mechanism by which heavy electrons break-up at the QCP, there are
%two physical analogies that may prove useful.
One of the challenges here is to understand the clear separation
between the single ion Kondo temperature $T_{K}$, which clearly
remains finite at the QCP, and the renormalized $T_{F}^{*} (B)$
which is driven continuously to zero. Burdin et al.\cite{burdin}
have suggested the analogy with pre-formed pairs in a
superconductor, according to which $T_{K}$ is the temperature at
which pre-formed local Kondo singlets develop and $T_{F}^{*}$ is
the scale at which phase coherence sets in to form mobile
quasiparticles.
 By analogy one can consider the QCP in heavy
fermions as a king of preformed scenario for the
 Kondo effect. In
such a picture, the presence of very strong antiferromagnetic
fluctuations close to a QCP dephases the heavy fermion composite
quasi-particles at increasingly low temperatures, as one gets
closer to the QCP. Eventually at the QCP the coherence temperature
is driven to zero, leading to the disintegration of the heavy
composite electron.

We would like to thank N. Andrei, J. Custers, P. Gegenwart
and  F. Steglich, for
for discussions related to this
work.  This research is  supported in  part by the  National
Science  Foundation grant NSF-DMR 9983156  (PC).

%
% The format of reference should be
% Author1, Author2, Author3, Journal {\bf volume} (year) page.
% No ``and'' between the authors are necessary.
%

\end{document}